\documentclass{article}
\usepackage{graphicx} 
\usepackage[utf8]{inputenc}
\usepackage[T1]{fontenc}
\usepackage{geometry}
\usepackage{setspace}
\usepackage{graphicx}
\usepackage{amsmath, amssymb, amsfonts}
\usepackage{physics}
\usepackage{bm}
\usepackage{tensor}
\usepackage{hyperref}
\usepackage{caption}
\usepackage{subcaption}
\usepackage{float}
\usepackage{titlesec}
\usepackage{tocloft}
\usepackage{fancyhdr}
\usepackage{xcolor}% Required for inserting images

\title{Quadratic Gauge Transformation}
\author{Akshit Sharma and Sunita Singh}

\begin{document}

\maketitle

\begin{abstract}
Symmetries play a significant role in understanding the conservation laws in quantum field theories. They are a result of the invariance of the system under certain set of transformations. Here, we attempted a dimensionless  quadratic  gauge transformation to achieve invariance in QFTs. We have explored these transformations to show the invariance of Abelian and Non- Abelian theories and established the conservation laws. We included an explicit graphical analysis to invoke the invariance. This is studied in a physical context, where different field configurations correspond to the same physical state. The necessity of the covariant derivative is studied in detail, highlighting how it ensures consistent transformation under local symmetry operations. We calculated the conserved current and charge explicitly and have shown their dependence on the quadratic gauge transformation. 
\end{abstract}

\section{Introduction}

Field theory plays a vital role in understanding the real world around us. Quantum Field Theory (QFT) provides the fundamental framework for describing physical systems in terms of fields and their interactions\cite{1,2}. A quanta is seen as the excitations of the fields, and their dynamics are guided by the laws of quantum mechanics and  special relativity. Along with these laws, symmetry plays a significant role in determining the structure and stability of physical theories. Symmetry, in a physical theory, describes the invariance of a system under formal transformations. These transformations may involve changes in space-time coordinates or internal properties of fields. An important result here is that the continuous symmetries are directly related with conserved quantities like energy, momentum,charge etc.These symmetries provide a direct connection between the formal structure and physical observables in a theory. This connection provides a systematic way to understand why certain quantities remain constant and how they are related to the underlying field dynamics\cite{3,4}. In the study of quantum fields, complex scalar fields are most important to study as they provide a simpler yet meaningful analysis for exploring symmetries. These fields exhibit not only  natural phase invariance but also result in the  emergence of conserved currents and charges. The analysis of such systems allows us to examine how field variables, their derivatives, and associated physical quantities are interconnected\cite{5}.

In this context, it becomes important to understand the role of local and global gauge transformation and their corresponding symmetries.  Here we would like to exploit higher order transformations particularly, local second order one. The focus is placed on analyzing key theoretical relations through explicit functional forms and graphical representation. Rather than treating graphical plots as supplementary illustrations, they are incorporated directly into the development of the theory. Each relation is first derived analytically, followed by the specification of field configurations that allow the relation to be expressed as a function of relevant variables. This approach makes it possible to study how physical quantities vary under different conditions and transformations.

These quadratic gauge transformations also represent an extension of the general local and global gauge transformations studied in the literature \cite{6,7,8}. They are also known by the name " Chirp phase" or " quadratic phase gauge"\cite{CF-1}. Such transformations are widely studied in the context of Landau level problem \cite{Landau1,Landau2,Landau3}, large scale transformations \cite{LST} and Non-commutative theories\cite{NC-1}. Further literature review,  shows the study of Non-Abelian\cite{NA1} and Chern-Simon theories\cite{NA2} in the context of the quadratic gauge transformations. The main theme of this paper is explore the quadratic gauge transformations for abelian and Non- Abelian theories explicitly through graphical demonstrations. The present literature does not emphasize on the formal structure conserved quantities associated with higher order transformation. The present study is an attempt to particularly analyze the formal structure of the Abelian and non-Abelian theories in coherent manner.    

\section{Quantum Fields and Gauge Invariance}
A quantum field theory is always consistent with the laws of quantum mechanics and the spatial theory of relativity \cite{1}. A field theory describes a quantum space filled with particles and their interaction with other particles. These particles come in the different categories depending on the type of constituting particles. On the basis of spin we can categories them as scalar, bosons and fermions. When we consider the formalism of the theory, it becomes relevant to discuss the invariant quantities like energy, momentum and other conserved currents. These conserved quatities are the result of some symmetric gauge transformations.We classify gauge transformations as global(gauge parameter is independent of space time) and local( gauge parameter depends on space-time variables). These two plays a key role in describing the symmetry in the theory. The global transformations are purely phase dependent and does not change the theory on the fundamental level.

\subsection{Local Gauge field Invariance}

 The local gauge transformations modifies the derivative in the formalism, and more interaction terms arise when we demand the invariance of the theory. Under global transformations, the symmetry properties of the complex scalar field shows that the phase of the field remains constant throughout space-time. This type of symmetry leads to the conservation of charge through Noether’s theorem. However, such transformations are restricted in the sense that they do not allow the phase to vary from point to point. To extend the analysis, we now consider local gauge transformations\cite{9,10}, in which the phase of the field is allowed to depend explicitly on space-time coordinates. The transformation is written as

\begin{equation}
    \begin{aligned}
        &\phi(x) \rightarrow \phi'(x) = e^{i\alpha(x)} \phi(x) \nonumber \\ 
        & \phi^{*}(x) \rightarrow {\phi^{*}}'(x) = e^{-i\alpha(x)} \phi^{*}(x) 
    \end{aligned}
\end{equation}

where $\alpha(x)$ is a real function of space-time. Demanding the gauge invariance of the Lagrangian, leads to symmetry breaking of the complex scalar field. To make theory invariant, we modify the derivation from partial to covariant. Using this definition, the covariant derivative acting on the scalar field is given by
\begin{equation}
D_\mu \phi(x) = \left( \partial_\mu + i e A_\mu(x) \right) \phi(x) \nonumber
\end{equation}

Here, $A_\mu(x)$ is a new field introduced into the theory, and $e$ is a constant parameter known as the coupling constant. The quantity $A_\mu(x)$ depends on space-time and will later be interpreted as a gauge field. The coupling constant $e$ determines the strength of the interaction between the scalar field and the gauge field.The field strength of the Abelian gauge field \cite{11,12,13,14}is given as 

\begin{equation}
  F_{\mu\nu} = \partial_\mu A_\nu - \partial_\nu A_\mu  
\end{equation}

Therefore, when we try to incorporate the gauge invariance of the combined Lagrangian density, the two fields transform as

\begin{equation}
    \begin{aligned}
       & D_\mu \phi(x) \rightarrow D'_\mu \phi'(x) = e^{i\alpha(x)} D_\mu \phi(x)  \nonumber \\
       & A_\mu(x) \rightarrow A'_\mu(x) = A_\mu(x) - \frac{1}{e} \partial_\mu \alpha(x)
    \end{aligned}
\end{equation}

The construction of the gauge invariant Lagrangian demonstrates that interactions can arise as a direct consequence of symmetry principles. The gauge field is not introduced arbitrarily; it emerges from the requirement that the theory remain invariant under local transformations. This result establishes a deep connection between symmetry and interaction. It shows that enforcing local gauge invariance leads to a consistent and physically meaningful description of interacting fields. The gauge invariant Lagrangian therefore forms the foundation for understanding interactions in quantum field theory. To determine the charge dependency of the U(1) gauge theory we rely mainly on the potential terms of the form. The last term, $e^2 A_\mu A^\mu \phi^* \phi$, represents the direct contribution of the gauge field to the energy density of the system. This term depends explicitly on the square of the gauge potential. To relate this to charge, we recall that physical quantities such as charge arise from the structure of the Lagrangian and the corresponding current. The presence of the term involving $A_\mu A^\mu$ indicates that the contribution to the charge depends on the magnitude of the gauge field. To obtain a functional form suitable for graphical analysis, we introduce a simplified field configuration. We assume that the scalar field amplitude is approximately constant in space-time, so that $\phi^* \phi = \text{constant} \nonumber$. Under this assumption, the dominant dependence on the gauge field arises from the term $A_\mu A^\mu$. For simplicity, we consider a configuration where only one component of the gauge field is significant and denote its magnitude by $A$. Thus, the contribution to the charge can be written as
\begin{equation}
Q(A) \propto A^2  \nonumber
\end{equation}

The graph shows a symmetric parabolic dependence of the charge on the gauge potential. The minimum value occurs at $A = 0$, and the charge increases with the magnitude of the gauge field. This behavior indicates that the contribution to the charge arises from the strength of the gauge field rather than its direction. The quadratic dependence reflects the structure of the interaction term in the Lagrangian, where the gauge field appears in squared form. Thus, the graphical representation provides a clear understanding of how the gauge field influences the conserved charge in the presence of interactions.

\begin{figure}[H]
    \centering
    \includegraphics[width=0.5\linewidth]{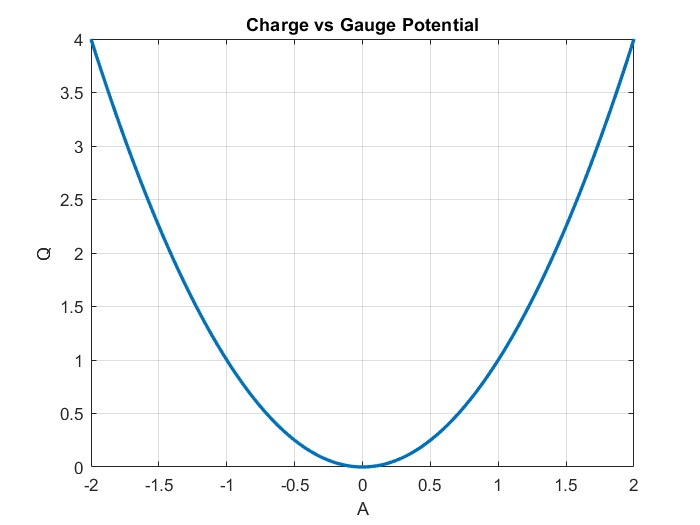}
    \caption{Gauge field Vs Charge}
    \label{fig: 2}
\end{figure}

The same graph is obtained when we keep the gauge field constant and the entire charge depends on the magnitude of the scalar field like $Q(A) \propto \phi^2  \nonumber$. Here also we observe that when the field becomes stronger, the corresponding charge increases. The quadratic dependence arises from the structure of the conserved current, where the charge is related to products of field components and their variations. The charge is always positive and grows in a parabolic manner.

\section{Quadratic (Gaussian) Gauge Transformation}
By studying the structure of the gauge transformations,we have realized that the gauge parameter plays an important role in discovering  the symmetries of a theory. In this section, we will try to explore the possible structure of it. Hence we propose a symmetric quadratic function as gauge parameter and we will try to  invoke it in the Lagrangian density function of various fields.

We take
\begin{equation}
\alpha(x) = \frac{x^2}{L^2}    \label{eq:GGT}
\end{equation}
where $L$ is a constant with the dimensions of length.

\subsection{Why Quadratic Transformation}

In this paper, we would like to explore the new quadratic phase transformations and check the invariance of the Abelian and Non-Abelian gauge theories in 4-dimension. Apriori,the considered quadratic looks like an obvious case of the generalized gauge transformations but they have meaningful physical significance. They represent U(1) local gauge transformation that grows without any bound and describing them somewhere between ordinary small gauge transformations and large topologically invariant ones. In the literature also such transformations are studied as the standard Fresnel diffraction integrals in optics, the Fourier integral operator theory in mathematics, and 'Bogoliubov transformation' in quantum optics \cite{Bogo1}. However, the literature does not provide an explicit study in the context of Abelian and Non-Abelian U(1) gauge theories. Our analysis is an attempt to fill this gap.

Another interesting feature of the considered quadratic gauge transformations is that for the Abelian gauge theories they correspond to the generalized Landau level problem \cite{Landau1,Landau2} for phase being $\frac{pi}{4}$. In case of Non- Abelian theories such transformations are important for  $W_\infty$ algebra in the lowest Landau level whose generators are quadratic \cite{Landau3,NA1}, in case QCD's for the large gauge transformation analysis the non trivial windings and vacuum angles are probed by such polynomial gauge transformations.\cite{T'hooft}. For String theory also the quadratic parameter just not represents the general multiplication but is governed by Courant brackets  describing a non-abelian structure that contains $exp(ix^{2})$-type elements as special cases, and whose failure to associate on coordinates is one of the  main structural properties.

We would like to explore their explicit relevance in Abelian and Non-Abelian theory. We attempt to understand their physical insights by plotting the conserved current against the transformation parameter. This article is an effort to understand more about the higher order gauge transformations and their consequences. 

\subsection{Abelian Gauge Theory and Quadratic Gauge Transformations}

In this section, we will try to explore the non-trivial nature of the gauge transformations and their application to Abelian Gauge theory (Quantum Electrodynamics)\cite{9,10} in 4- dimensions. The Lagrangian density of the same is given as 

\begin{equation}
\mathcal{L} = (D_\mu \phi)^* (D^\mu \phi) - m^2 \phi^* \phi-\frac{1}{4}{F_{\mu\nu}}{F^{\mu\nu}}    \nonumber
\end{equation}  

Here 
\begin{equation}
F_{\mu\nu} = D_\mu A_\nu - D_\nu A_\mu   \nonumber
\end{equation}

The covariant derivative is defined as
\begin{equation}
D_\mu \phi(x) = \left( \partial_\mu + i e A_\mu(x) \right) \phi(x)
\end{equation}

The above Lagrangian density describes the interaction of matter with photons.
Let us now examine the invariance under the quadratic transformations term by term. The gauge function that we have used is defined above and the derivative of which can be defined as follow
\begin{equation}
\partial_\mu U(x)= \partial_\mu \alpha(x)=2\frac{\partial_\mu x}{L^2} \nonumber
\end{equation}

So after analyzing the operation of the covariant derivation on the transformed scalar matter field we observe that it remains constant 

\begin{equation}
   \begin{aligned}
D'_\mu \phi'(x) = e^{i\alpha(x)} D_\mu \phi(x) \\
\phi'(x) \phi'^{*}(x)=\phi(x) \phi^{*}(x)    \nonumber
   \end{aligned} 
\end{equation}

 Now, we start with the kinetic energy term of the matter field,as shown above the covariant derivative is invariant under the transformations and overall phase also gets cancel, we conclude that the matter field remains invariant. Further consider the massless U(1) gauge theory, the kinetic energy terms is explored 

\begin{equation}
  F'_{\mu\nu}= \left( \partial_\mu + i e A'_\mu(x) \right) A'_\nu - \left( \partial_\nu + i e A'_\nu(x) \right)A'_\mu(x) \nonumber
\end{equation}
where we have considered that

\begin{equation}
A(x) \rightarrow A'(x) = A(x) + \frac{1}{e} \frac{2x}{L^2}\nonumber
\end{equation}

Here the gauge potential is not unique but varies with respect to the space time co-ordinates. This feature of the gauge transformations also matches the ambiguity in the measurement of the potential.Conclusively, the potential is never unique whereas the field strength is the fundamental unique measurable quantity. After calculations we can show

\begin{equation}
    F'_{\mu\nu}= F_{\mu\nu}-\frac{2}{e}\left(\partial_{\mu} x_{\nu} -\partial_\nu x_\mu \right) \nonumber
\end{equation}

\begin{figure}
     \centering
     \begin{subfigure}[b]{0.495\textwidth}
         \centering
         \includegraphics[width=\textwidth]{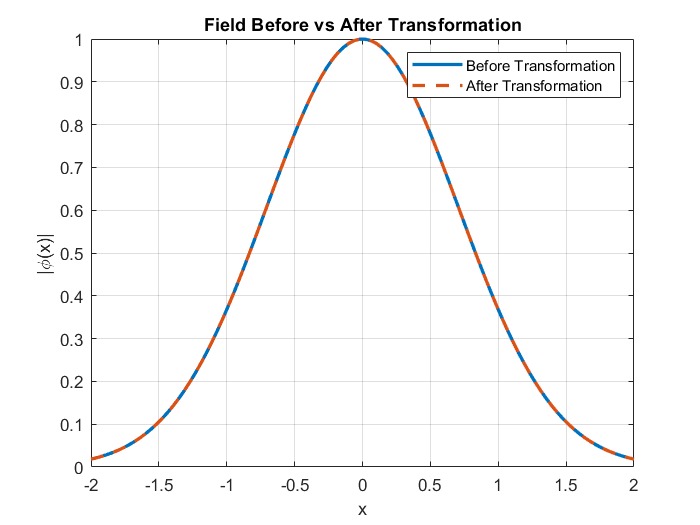}
         \caption{Complex Scalar Field Vs Transformations}
         \label{fig:3(a)}
     \end{subfigure}
     \hfill
     \begin{subfigure}[b]{0.495\textwidth}
         \centering
         \includegraphics[width=\textwidth]{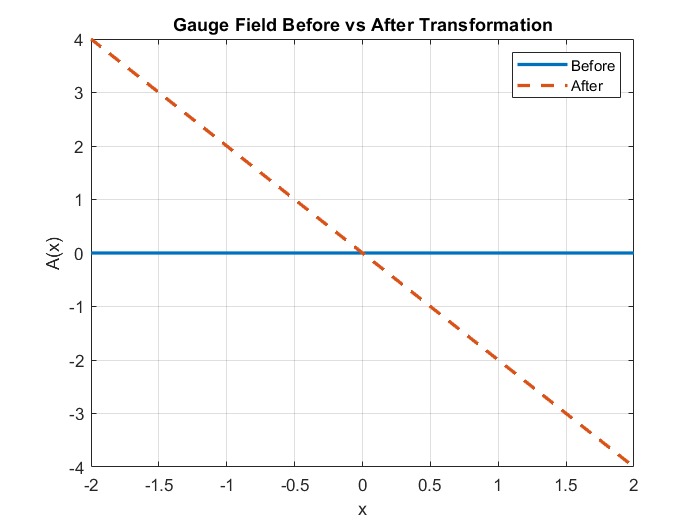}
         \caption{Gauge Field Vs Transformations}
         \label{fig:3(b)}
     \end{subfigure}
     \label{fig:3}
\end{figure}

As we are working in a flat spacetime with metric $\eta_{\mu \nu}$ it is easy to show that the last term of the above equation is zero. Hence we can conclude that gauge field strength is also invariant under the quadratic gauge transformations transformations. However, for $\alpha=constant$, $A_{\mu}$ will not change but that's a trivial choice. The quadratic transformation $\alpha= x^{\mu}x_{\mu}$, $A_{\mu}$ grows linearly and represents an unbounded function. So,it becomes mandatory to see its relevance graphically in the figure {\ref{fig:3(a)}}. This behavior reflects the derivative of the gauge function, which introduces a position-dependent shift.  An interesting feature is that the complex scalar field remains invariant before and after the transformation operation but the gauge potential attains a linear dependency. Thus, the graphical analysis demonstrates that gauge transformations can modify the gauge field configuration without altering the underlying physical observables.

\subsection{Conserved Current density}
As the above considered transformations doesn't change the formal structure of the theory we now calculate the conserved charges in the formalism. Hence, as per Noether's theorem the conserved charge density and Noether's  current is given by 

\begin{equation}
   \begin{aligned}
      J^{\mu} &= \sum_{i} \frac{\partial L} {\partial(\partial_{\mu}\phi)} \delta(\phi_{i})  \nonumber \\
      J^{\mu} &= \frac{\partial L} {\partial(\partial_{\mu}\phi)} \delta \phi+ \frac{\partial L} {\partial(\partial_{\mu}\phi*)} \delta \phi^*+ \frac{\partial L} {\partial(\partial_{\mu} A_\nu)} \delta A_{\nu}
   \end{aligned}
\end{equation}
Calculating the contribution term wise, we get
\begin{equation}
    \begin{aligned}
        \frac{\partial L} {\partial(\partial_{\mu}\phi)} &= (D^\mu \phi)^{*}\\
        \frac{\partial L} {\partial(\partial_{\mu}\phi^*)} &= (D^\mu \phi) \nonumber \\
        \frac{\partial L} {\partial(\partial_{\mu} A_\nu)} &= F^{\mu\nu}
    \end{aligned}
\end{equation}

Now the net charge density is
\begin{equation}
J^{\mu}= ix^2[(D^\mu \phi)^{*}\phi- (D^\mu \phi) \phi^*]-\frac{2}{e}F^{\mu\nu}x_\nu 
\end{equation}
and the corresponding charge density is defined as 
\begin{equation}
    J^\mu= {j^\mu} _{Scalar} -\frac{2}{e}F^{\mu\nu} x_\nu  \nonumber
\end{equation}

Apriori, the transformed charge density looks space-time dependent and varies as the gauge parameter changes but its conservation is important for the invariance theory.This tells us that the non trivial behavior of the current in the theory. Moreover, for the particular case, the transformations is independent of the time derivative. Therefore, the charge remains invariant. We can show term wise conservation or collective invariance. Hence, the on-shell condition is $\partial_\mu J^\mu =0$.

\begin{equation}
    \begin{aligned}
         \partial_\mu J^\mu &= \partial_\mu \left(x^2 {j^\mu} _{Scalar}  -\frac{2}{e}F^{\mu\nu} x_\nu \right) \nonumber \\
         &= 2x {j^\mu} _{Scalar}-x^2 \partial_\mu {j^\mu} _{Scalar} - \frac{2}{e}\partial_\mu (F^{\mu\nu}) x_\nu - \frac{2}{e}F^{\mu\nu} \partial_\mu x_\nu   =0     
    \end{aligned}
\end{equation}

where we used Maxwell's equation to prove the invariance of the combined current density under quadratic gauge transformations. For calculating the conserved charge we need only the time component of the conserved current density

\begin{equation}
  \begin{aligned}
     Q &=\int d^3 x  J^0 (x) \\
      J^0&= ix^2[(D^0 \phi)^{*}\phi- (D^0 \phi) \phi^*]-\frac{2}{e}F^{0\nu}x_\nu  \nonumber \\
\implies Q &= \int d^3 x \left( x^2[ie( \partial ^0 \phi^*\phi- \partial^0 \phi \phi^*) +2 e^2 A_0|\phi|^2]-\frac{2}{e} \boldsymbol{E}.\boldsymbol{x} \right)      
  \end{aligned}
\end{equation}

The charge density corresponding to the scalar field gets scaled by a factor of $x^2$ and the current also becomes position-dependent due to the spatial variation of the gauge function. The total charge is the algebraic sum of the scalar part and the gauge part which can be shown to remain invariant if we use Maxwell's equation $\boldsymbol{\nabla}.\boldsymbol{E} =e j^0_{scalar}$. Here we may conclude that at first the charge possess space dependency but it can easily be shown that it is conserved in time.  Further, the induced current is zero at the origin and increases with distance, reflecting the gradient of the phase. 

Interesting features are seen when we plot a graph between the $j^0_{scalarl}$, $j_{gauge}$, $J_{total}$ versus the gauge transformations. We tried to take the gaussian trend with $\sigma= 0.7$, $\lambda=\frac{1}{L^2}= 1.7$,$q=1.1$,$ E_0= 0.7$ and $w=1$ 

\begin{figure}
    \centering
    \includegraphics[width=0.5\linewidth]{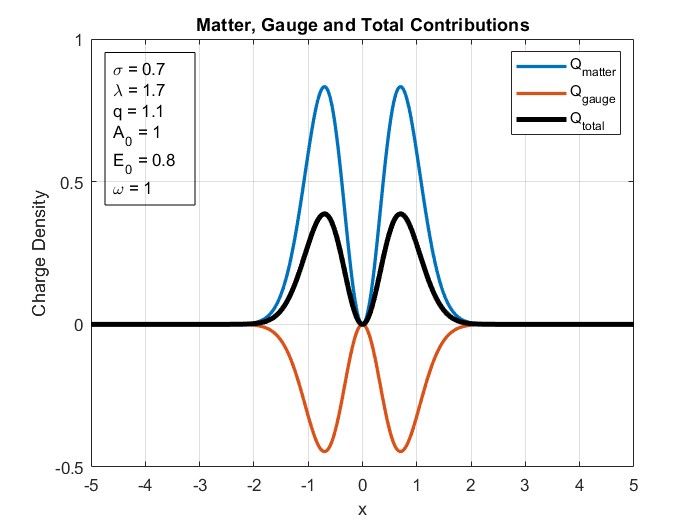}
    \caption{Conserved Charge versus transformation }
    \label{fig:4}
\end{figure}

\begin{itemize}
    \item For ordinary matter the charge takes a maximum value at $x=0$ but here the charge gets weighted by a factor of $x^2$. Therefore, the matter charge is zero at the center and peaks $x=\pm 1$ value describing the radius of charge distribution.
    \item The second term in the charge expression is from the gauge part, the main contribution is of electric field and linear space co-ordinate. The overall pattern is similar the scalar field part with a negative peak of charge which decays to zero at $x=0$.
    \item As we concluded above that the total charge is the combination of the scalar (matter) and the gauge part. Though the two peaks up in completely opposite direction, magnitude of the matter charge is always higher in comparison to the gauge contribution. Therefore, the net charge is positive, peaks up at $x=\pm 1$ and decreases to zero value at $x=0$.
    \item As the total Noether charge is proportional to gauge parameter it increases with quadratically and shows the spatial arrangement of charge density.
    \item Conclusively, we can say that the new charge is not discrete but refers to a spatially spread region describing the availability of particles.It describes a double peak structure centered at $x=0$.
\end{itemize}

\subsection{Non Abelian Gauge Fields}
After analyzing the charge in Abelian case, we can now look for the same in Non-Abelian gauge theories. A comparative study provides a new interpretation for the scaled Noether's charge. In the Abelian case, a single gauge field $A_\mu(x)$ was introduced corresponding to the U(1) gauge group . However, in the non-Abelian case, the symmetry group has multiple generators, and each generator is associated with a corresponding gauge field and gauge group $SU(N)$ \cite{11,12}.

Thus, the gauge field becomes a set of fields, written as
\begin{equation}
A_\mu(x) = A_\mu^a(x) T^a     \nonumber
\end{equation}
where $A_\mu^a(x)$ are the components of the gauge field, and $T^a$ are the generators of the symmetry group $SU(N)$.The covariant derivative is generalized as
\begin{equation}
D_\mu \phi = \partial_\mu \phi + i g A_\mu^a(x) T^a \phi   \nonumber
\end{equation}

Here, $g$ is the coupling constant that determines the strength of the gauge interaction. The field strength tensor is defined through the commutator of covariant derivatives.The covariant derivative in the theory also includes the Lie generators and therefore gives rise to a new non-commutative term responsible for a Chiron-Simon theory. The field strength tensor becomes
\begin{equation}
F_{\mu\nu} = \partial_\mu A_\nu - \partial_\nu A_\mu + i g [A_\mu, A_\nu]   \nonumber 
\end{equation}

Substituting $A_\mu = A_\mu^a T^a$, where $T^a$ is the generator of the $SU(N)$ group that follow the Lie algebra as 
\begin{equation}
    [T^a,T^b]=if^{abc} T^c ,   Tr(T^a T^b)=\frac{1}{2} \delta ^{ab} \nonumber
\end{equation}

The complete Lagrangian density for the Yang- Mills theory is given as
\begin{equation}
    \mathcal{L}= -\frac{1}{2} Tr(F_{\mu\nu}F^{\mu\nu})+(D_\mu \phi)^{\dagger}(D^\mu \phi)-m^2 \phi^{\dagger} \phi   \nonumber
\end{equation}

Now, operating the quadratic transformations the field transforms like

\begin{equation}
   \begin{aligned}
       \phi(x) &\rightarrow U(x) \phi(x) \\
     A_\mu(x) &\rightarrow A'_\mu(x) = U(x) A_\mu(x) U^{-1}(x) - \frac{i}{g} (\partial_\mu U(x)) U^{-1}(x) \nonumber
        \end{aligned}
\end{equation}

This transformation is more complex than the Abelian case due to the matrix structure of $SU(N)$. The introduction of Yang--Mills fields generalizes the concept of gauge invariance to systems with multiple interacting components. The presence of multiple gauge fields and their transformation properties leads to new interaction terms that are not present in Abelian theory. These features form the foundation of non-Abelian gauge theories, where the interaction structure is determined by the symmetry group itself\cite{16,17,18}. For implementing the quadratic gauge transformations, we must consider the gauge group structure of the gauge parameters  as 
\begin{equation}
    \begin{aligned}
       & U(x) =exp(i\alpha^a(x))=exp(i\frac{x^\mu x_{\mu}}{L^2}T^a) \nonumber   \\
       & \partial_\mu(\alpha^a(x))= 2\frac{x_\mu}{L^2}  \nonumber \\
       & U\approx 1+i\frac{x^2}{L^2}T^a
    \end{aligned}
\end{equation}

The last expression shows the infinitesimal transformations for x being a small. Let us now check the term wise variation of the Lagrangian density of Yang- Mills theory. First consider the matter field as 

\begin{equation}
    \begin{aligned}
        &\phi \implies \phi' = U\phi=  exp(i\frac{x^\mu x_{\mu}}{L^2}T^a) \phi  \\
        &\delta \phi =i\frac{x^\mu x_{\mu}}{L^2}T^a \phi  \\
        &\partial_\mu \phi'= U(x) (\partial_\mu \phi + 2i\frac{ x_{\mu}}{L^2}T^a \phi)  \\
        & A_\mu \implies A\mu ' =UAU^\dagger+\frac{i}{g}(\partial_\mu U) U^\dagger          \\
        &\delta A_\mu = -\frac{x^\nu x_\nu}{L^2} f^{abc} A_\mu +2 \frac{x^\mu}{gL^2} T^a   \\
        &(D_\mu \phi)' \implies  D_\mu' \phi' = (\partial_\mu -ig A_\mu')\phi' = U(D_\mu \phi)\\
        & F_{\mu \nu}'= U F_{\mu\nu}U^\dagger    \nonumber 
    \end{aligned}    
\end{equation}

The extra term in matter field signifies a pure non-Abelian contribution and gets cancel with gauge field transformation part. Interestingly, the covariant derivative transforms exactly in the fundamental representation confirming the absorption of additional terms. After analyzing term wise invariance we are now collecting all terms to prove the invariance of the Lagrangian. As already shown in the transformations above, all the terms of the Lagrangian density transform in a covariant manner, and the new terms arising due to the quadratic gauge used in the paper are canceled by their counterparts. As a result, the Lagrangian density is completely invariant under the new type of local gaussian gauge transformations used in the paper. We are now apply the Noether's theorem and calculate the conserved current in the theory. Now using the expression of conserved current density as
\begin{equation}
          J^{\mu} = \sum_{i} \frac{\partial L} {\partial(\partial_{\mu}\phi)} \delta(\phi_{i})
\end{equation}

We can expand the Noether's current as the contribution coming from the matter part and the gauge part. For matter part, we obtain the expression

\begin{equation}
    J^{\mu a}_{matter} = \frac{x^2}{L^2}\left[i(\partial^\mu\phi^\dagger )T^a \phi -i\phi^\dagger T^a \partial^\mu\phi +igf^{abc} A^{b} \phi^\dagger T   \right]
\end{equation}

The calculated matter current turns out to be scaled by a gauge parameter $\alpha(x)=\frac{x^2}{L^2}$ to the original Yang-Mills current. This scaling indicates that the conserved current is now a space dependent quantity and varies quadratically. For the gauge part also, we calculate the contributions as 

\begin{equation}
    J^{\mu a}_{gauge}= -\frac{2}{gL^2}F^{\mu\nu a}x_\nu-\frac{x^2}{L^2}f^{abc}A^b_\nu F^{\mu\nu c} 
\end{equation}

 The combined conserved current expression is
\begin{equation}
     J^{\mu a} = \frac{x^2}{L^2}\left[i(\partial^\mu\phi^\dagger )T^a \phi -i\phi^\dagger T^a \partial^\mu\phi +igf^{abc} A^{b} \phi^\dagger T   \right] -\frac{2}{gL^2}F^{\mu\nu a}x_\nu-\frac{x^2}{L^2}f^{abc}A^b_\nu F^{\mu\nu c} \nonumber   
\end{equation}

\begin{equation}
    J^{\mu a} = J^{\mu a}_{YM} -\frac{2}{gL^2}F^{\mu\nu a}x_\nu-\frac{x^2}{L^2}f^{abc}A^b_\nu F^{\mu\nu c} \label{YM Current}
\end{equation}

After equating the above expression with the standard Yang-Mills conserved current we observe that the matter part matches with Yang-Mills theory scaled by a factor of $\alpha(x)=\frac{x^2}{L^2}$. The second term is the pure gauge contribution and shows linear variation with the space co-ordinates. Moreover, the last term is the non-Abelian contribution which is also scaled by the gauge parameter and shows quadratic variation. 

\begin{figure}
    \centering
    \includegraphics[width=0.5\linewidth]{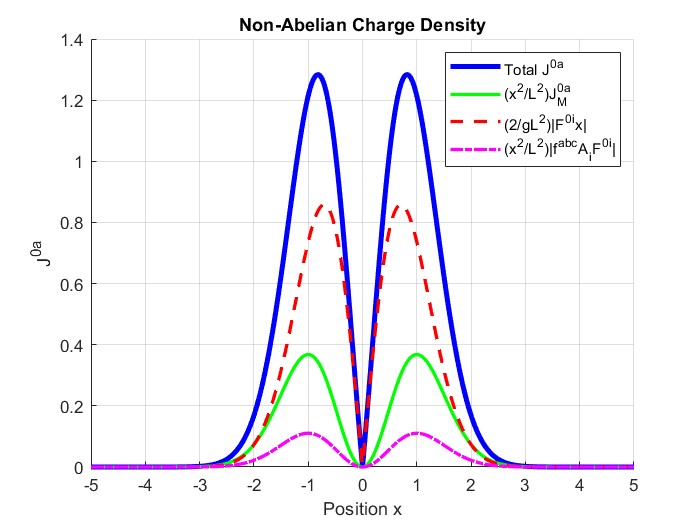}
    \caption{Variation of Non-Abelian Current with transformations}
    \label{fig:5}
\end{figure}
The graph obtained between the current and the space parameter shows similar features of the Abelian case. The graph is drawn for group $SU(2)$ by taking values $\sigma=1,g=1$ and $\frac{1}{L^2}=1$. Importantly, the height and position of the peaks is now determine by the color coupling. The gauge contribution in the second term is linear and grows as the space parameter increases.The last term is the result of the commutator relation of gauge fields. It has even parity and varies symmetrically with gauge parameter. For small 'L' values, the coupling term dominates over the matter part. We can now compare the features of Abelian and Non-Abelian charges in a tabular manner like

\begin{table}
    \centering
    \begin{tabular}{|c|c|c|} \\  \hline
       \textbf{Property}  & \textbf{Abelian Case} & \textbf{Non-Abelain} \\  \hline
       Number of Charge  & One Charge  & \  3 For SU(2) and 8 for SU(3) \\ \hline
        Commutator term & Always Zero &  Always Non-zero \\ \hline   
        Variation with $\frac{1}{L^2}$  & Linear   & Quadratic  \\   \hline
        Conservation Law  & $\partial_\mu J^\mu$   &  $(D_\mu J^\mu)^a$   \\   \hline
        
    \end{tabular}
    \caption{Comparison between Abelian and Non- Abelian Theory}
    \label{Table:1}
\end{table}

We can also look for exact way of proving the above current and the covariant equation turns out to be  

\begin{equation}
  \begin{aligned}
    &(D_\mu J^\mu)^a=0   \label{Current Cons} \\
    &= \partial_\mu J^{\mu a} +gf^{abc} A^b_\mu J^{\mu a}=0   \\
  \end{aligned}
\end{equation}

we can prove the conservation of current of individual contribution of the matter part and the gauge part from equation (\ref{YM Current}). First, consider the matter part where we can apply covariant conservation of current and show that  as  
\begin{equation}
    \begin{aligned}
        & \partial_\mu(\frac{x^2}{L^2}J^{\mu a}_{YM})=2\frac{x_\mu}{L^2} J^{\mu a}_{YM}+\frac{x^2}{L^2}\partial_\mu J^{\mu a}_{YM}  \nonumber \\
        & \partial_\mu J^{\mu a} _{YM}=-gf^{abc} A^b_\mu J^{\mu a}\\    
        \implies  & \partial_\mu(\frac{x^2}{L^2}J^{\mu a}_{YM})=2\frac{x_\mu}{L^2} J^{\mu a}_{YM}-\frac{x^2 g}{L^2}f^{abc} A^b_\mu J^{\mu a}     \label{matter part}\\
        &\partial_\mu (\frac{2}{gL^2}F^{\mu\nu a}x_\nu)= -2\frac{x_\mu}{L^2} J^{\mu a} + \frac{x_\nu}{L^2} f^{abc}A^b_\mu F^{\mu\nu} 
    \end{aligned}
\end{equation}

Now after combining the matter part and gauge part variations of the current we observe that the last term in equation (\ref{YM Current}) gets cancel and the overall charge is conserved as assumed in equation (\ref{Current Cons}). Further the expression for conserved charge in the transformed Yang-Mills theory is given by \cite{19} 

\begin{equation}
{Q}=-\frac{1}{L^2} \int dx^3  \left[ x^2 J^0_{YM}(x)-\frac{2}{g}E^a(x).x -x^2 f^{abc} A^b_i(x)E^{ic}(x)    \right]
\end{equation}

The presence of the commutator term is the key feature of non-Abelian gauge theory. It shows that the gauge fields themselves interact with each other.
This is in contrast to the Abelian case, where the gauge field does not self-interact. The non-zero commutator arises from the non-commuting nature of the generators. Thus, the field strength tensor not only describes how the field varies but also includes contributions from the interaction between gauge fields. The additional term in the field strength tensor leads to new interaction effects that are absent in Abelian theory. It is this term that gives rise to the self-interaction of gauge fields in Yang--Mills theory.

\section{Conclusion}

We understand the important role of invariance and symmetries through gauge theory invariance. Here, we presented a critical analysis of local gauge invariance in the complex scalar, Abelian and Non-Abelian. Each framework provides a different level of complexity in terms of symmetry and interaction. Symmetry plays a central role in determining the structure of physical theories. The transition from global to local symmetry introduces significant changes in the behavior of fields. In the considered local gauge symmetry, we obtain extra terms in the derivative. This leads to modification of partial derivative to covariant derivative in each theory. This leads to a new type of interaction between the scalar field and the gauge field. In Non-Abelian theory, the symmetry group is $SU(N)$, and have an additional self interaction terms of the form $A\mu A\nu$. This feature is intrinsic to the theory due to its non-commutative nature.

It is observed that in both theories the overall current gets scaled by space co-ordinate. The matter part of the Lagrangian density is scaled by quadratic factor and the gauge part has a linear dependence. Despite the new factors, the overall current remains conserved in the theory.  The total charge is the algebraic sum of the scalar part and the gauge part which can be shown to remain invariant if we use Maxwell's equation $\boldsymbol{\nabla}.\boldsymbol{E} =e j^0_{scalar}$. Here we may conclude that at first the charge possess space dependency but it can easily be shown that it is conserved in time.  Further, the induced current is zero at the origin and increases with distance, reflecting the gradient of the phase.       

An important aspect of the study is to show explicit graphical study including complex scalar fields, Abelian and Non- Abelian gauge Lagrangian density as a function of charge and space parameters. Interestingly, after the quadratic gauge transformations we found, they modify the gauge field configuration without altering the underlying physical observables. From the analysis, we also infer that the charge density remains unchanged, confirming invariance of total charge. Graph shows a double peak structure, due to quadratic scaling. It shows a symmetric distribution about the origin and and is always zero beyond certain range of the space co-ordinate. The contribution of the gauge part is opposite to the matter part and decreases the over all charge in the theory. Similar features are observedfor the Non-Abelian charge as well.    

The formulation and analysis of the quadratic transformation we have shown that local quantities may vary, global quantities in particular total charge remain conserved. In non-Abelian theory, the field strength tensor turns out to be invariant for a Minkowiski type space time. The theory also includes additional terms that are self-interacting of gauge fields. The comparison between Abelian and non-Abelian cases shows that non-Abelian fields exhibit non-commutative behavior due to interaction among gauge fields. 

\section{Acknowledgment and Funding}
The authors acknowledge the Department of Physics, Kirori Mal College,
University of Delhi, and thank the Dr~N.~S.~Pradhan Memorial Library. This project is not funded by any funding agency.

\end{document}